\documentclass[3p,times]{elsarticle}

\usepackage{amssymb}

\usepackage[figuresright]{rotating}

\usepackage{amssymb} 
\usepackage{amsmath} 
\usepackage{algorithm}
\usepackage{algpseudocode}
\usepackage{tikz}
\usepackage{makecell}
\usepackage[final]{listings}
\usepackage[T1]{fontenc} 
\usepackage[hidelinks]{hyperref}

\usetikzlibrary{patterns}
\definecolor{myellow}{RGB}{255, 193, 7}
\definecolor{myblue}{RGB}{30, 136, 229}

\def\bm#1{\mbox{\boldmath{$#1$}}}

\begin{document}

\begin{frontmatter}



  \title{Efficient Extraction of Atomization Processes from High-Fidelity Simulations}
  
  \author{Brendan Christensen \corref{cor1}}
  
  \ead{brendan.christensen@montana.edu}
  
  \author{Mark Owkes}

  \cortext[cor1]{Corresponding Author}


\address{Department of Mechanical and Industrial Engineering, Montana State University, Bozeman, MT, 59717-3800, USA}

\begin{abstract}
Understanding the process of primary and secondary atomization in liquid jets is crucial in describing spray distribution and droplet geometry for industrial applications and is essential in the development of physics-based low-fidelity atomization models that can quickly predict these sprays. Significant advances in numerical modelling and computational resources allows research groups to conduct detailed numerical simulations and accurately predict the physics of atomization. These simulations can produce hundreds of terabytes of data. The substantial size of these data sets limits researchers’ ability to analyze them. Consequently, the process of a coherent liquid core breaking into droplets has not been analyzed in simulation results even though a complete description of the jet dynamics exists. The present work applies a droplet physics extraction technique to high-fidelity simulations to track breakup events and data associated with the local flow. The data on the atomization process are stored in a Neo4j graphical database providing an easily accessible format. Results provide a robust, quantitative description of the process of atomization and the details on the local flow field will be useful in the development of low-fidelity atomization models.  
\end{abstract}

\end{frontmatter}

\section{Introduction}
Atomizing sprays have a wide range of applications in industrial and environmental fields (e.g. fuel injection~\cite{Zhao1999}, agricultural sprays~\cite{Law2001}, pharmaceutical sprays~\cite{Maa1999}, and air-sea interaction~\cite{Fuentes2010}). Consequently, the body of research on this topic is vast and spans multiple centuries~\cite{Savart1833,Rayleigh1878}. Researchers have made significant progress in understanding many aspects of the atomization process, however, the mechanisms that drive instabilities in coherent liquid structures and ultimately cause breakup are still widely unresolved. This is, in part, due to limitations in experimental data collection from atomizing flows. Visual analysis of atomizing jets is difficult because of the scale and speed at which these systems develop. Furthermore, the development of droplets creates an opaque cloud, which blocks view of the liquid core and severely limits the ability to study primary instability and breakup mechanics. Recent advances in ballistic~\cite{Linne2005,Sedarsky2010,Slangen2016} and x-ray~\cite{Wang2006,Heindel2018,Li2021} imaging of atomizing flows have attempted to remedy visualization issues. These methods provide snapshots of the atomization process and lack important temporal information. Because of challenges with current experimental methodology, numerical simulations, and particularly high-fidelity simulations that resolve the relevant time and length scales, have been developed and provide an alternative method to study the physics of atomization.  

Advances in numerical methods and computational efficiency within the past decade have greatly expanded the capabilities of numerical simulations. Researchers are now able to simulate multiphase flows with exceptional accuracy. Numerically simulating multiple phases is not a trivial process, however. These systems involve a wide range of topological scales in fluid and turbulent structures in addition to jumps in physical qualities at the phase interfaces. Significant effort has been and is still invested in improving numerical simulation methodology to more accurately and efficiently simulate atomizing systems. Accordingly, most high-fidelity simulation studies on atomization to date are focused on the development of numerical methods and testing against experimental results rather than using the simulations to produce novel atomization data. But, with the pending maturity of the field, several research groups have been able to gather some insightful statistics from high-fidelity atomization simulations. 

Following Gorokhovski and Herrmann's review of atomization simulations~\cite{Herrmann2008}, the field developed rapidly. Among the first high-fidelity simulations of atomizing liquid were Shinjo and Umemura~\cite{Shinjo2010} and Desjardins and Pitsch~\cite{Desjardins2010} who simulated round and planar liquid jets, respectively. These works provided a foundation for future simulation studies by identifying numerous breakup mechanisms and providing qualitative descriptions of previously unclassified processes. Following these pioneering studies, various other works were published which utilized high-fidelity simulations to expand upon existing descriptions and identify novel breakup mechanisms. Several of these groups incorporated vorticity measurements into their post-processing to strengthen descriptions of breakup by integrating this additional dimension to their analysis \cite{Shinjo2011b,Jarrahbashi2014,Watanabe2014,Jarrahbashi2016,Li2016,Zandian2017,Zandian2018}. Additionally, many researchers focus their efforts on comparing linear-stability theory~\cite{Reitz1987,Beale1999} and high-fidelity simulations~\cite{Shinjo2011b,Fuster2013,Deshpande2015,Trujillo2018,Krolick2018}. These studies process visualizations of the atomizing jet to analyze and track wavelengths as they become more unstable, ultimately leading to breakup. Several recent studies have obtained turbulence statistics from throughout the simulation using constrained spatial and/or temporal domains to limit data-processing requirements~\cite{Hasslberger2019,torregrosa2020,Salvador2018}. 

The numerical studies mentioned above have greatly improved understanding of the atomization process. They developed robust qualitative descriptions of breakup events and utilized bulk quantities to elucidate some of the underlying physics of liquid breakup. However, despite the recent advances in atomization simulations, there remain significant gaps in our understanding of the process. This is primarily due to a lack of available methodology for obtaining many relevant statistics from high-fidelity simulations. In particular, current research methods have not been able to extract local, temporally continuous, quantitative data on liquid breakup. The data include information on droplet shape and size characteristics, local flow field data in both the liquid and gas phases, and information on how these values evolve temporally and spatially throughout a spray system. An appreciable challenge obtaining this data is derived from the massive size of resultant data sets; these can be hundreds of terabytes or even larger. Parsing such data sets for relevant information is not practical and is often impossible.   

While numerical simulations are becoming more efficient, it remains computationally expensive and requires the use of high-performance computing. Reduced-order atomization models aim to provide a viable alternative to high-fidelity simulations at a fraction of the computational cost. Some prominent examples of atomization models include the Taylor analogy breakup (TAB) model \cite{TAB1987}, the Pilch-Erdman (PE) model \cite{Pilch1987}, the Kelvin-Helmholtz-Rayleigh-Taylor (KH-RT) model \cite{Reitz1987,Teclllwlvgy1987,Patterson1998}, the Eulerian-Lagrangian spray and atomization (ELSA) model \cite{Wang2011}, the bag type breakup (BTB) model \cite{Wang2014}, the multi-mode breakup (MMB) model \cite{Wang2015}, and the modified Taylor analogy breakup (MTAB) model \cite{Sula2020}. These atomization models try to predict how and when a liquid structure breaks apart without fully resolving the physics. As mentioned above, however, there is little information on the mechanisms that control breakup events and even less information on the process of atomization. Model developers would greatly benefit from local data sampled from breakup events throughout the simulation.

The present work aims to address the above obstacles through the improvement of a framework, first introduced in Rubel and Owkes~\cite{RubelOwkes2019}, which extracts relevant data from breakup and coalescence events in atomization simulations. Through the use of this tool, many researchers would gain access to extensive and easily accessible data sets from atomization simulations. Using this data, they could build on previous work, introduce high-quality statistical analyses, and significantly advance the field. Additionally, the structure geometry and flow field can be assessed to quantify the details of the breakup conditions. This new information will advance our understanding of the atomization process and provide statistics with which to inform and improve reduced-order atomization models.

\section {Methods}
The droplet physics extraction tool, originally proposed in Rubel and Owkes \cite{RubelOwkes2019}, is improved and applied in this work. The tool is implemented into high-fidelity atomization simulations to gather data from discrete events within these simulations coinciding to liquid breakup and coalescence. This section will outline the methodology for this process in detail. 

\subsection{Computational Platform}
The proposed extraction tool can be applied to any high-fidelity Navier-Stokes solver, given the solver incorporates two identification numbers used to determine breakup and coalescence events, which will be defined in the following sections. This work employs the NGA computational platform \cite{Desjardins2008a,Desjardins2008b,Owkes2014,Owkes2017}. NGA solves the two-phase formulations of the Navier-Stokes mass and momentum conservation equations, defined as 
\begin{equation}
    \frac{\partial \rho_{\phi}}{\partial t}+\nabla\cdot(\rho_{\phi}\bm{u}_{\phi})=0
\end{equation}
and 
\begin{equation}
    \frac{\partial \rho_{\phi} \bm{u}_{\phi}}{\partial t}+\nabla\cdot(\rho_{\phi}\bm{u}_{\phi}\otimes \bm{u}_{\phi})= -\nabla p_{\phi} +\nabla\cdot(\mu_{\phi}[\nabla \bm{u}_{\phi} +\nabla \bm{u}^{T}_{\phi}]) +\rho_{\phi}\bm{g}
\end{equation}
respectively. In these equations, $\rho_{\phi}$ is density, $\bm{u}_{\phi}$ is the velocity vector, $p_{\phi}$ is the pressure, $\mu_{\phi}$ is the dynamic viscosity, $t$ is time, $\bm{g}$ is the gravitational acceleration vector and $\phi$ is the phase indicator with either $\phi = g$ (gas) or $\phi = l$ (liquid). These equations are solved on a staggered Cartesian mesh, in which scalar values, such as pressure, are stored at cell centers and velocities at the cell faces. Time is discretized with an iterative Crank-Nicolson formulation and a semi-implicit correction is applied on each sub-iteration \cite{Choi1994}. Away from the phase interface, mass, momentum, and any other scalars are transported with conservative, high-order finite difference operators~\cite{Desjardins2008a}. Near the interface, an un-split semi-Lagrangian geometric volume-of-fluid (VOF) method is utilized to ensure conservative transport of mass and momentum \cite{Owkes2014,Owkes2017}. The interface is reconstructed using a piece-wise linear interface reconstruction (PLIC) \cite{Youngs1982} with interface normal vectors computed  using the efficient least-squares VOF interface reconstruction algorithm (ELVIRA) \cite{Pilliod2004}. The pressure Poisson equation is solved utilizing the ghost fluid method \cite{Fedkiw1999} and a black box solver \cite{Dendy1982}. Interface curvature is computed with the adjustable curvature evaluation scale (ACES) method \cite{Owkes2018}. NGA is fully parallelized with message passing interface (MPI) and scales well to tens-of-thousands of cores~\cite{Desjardins2010}. 

\subsection{Breakup and Coalescence Event Identification}
The coalescence and split identification processes operate through the implementation of two identification numbers, which are integers unique to every independent liquid structure within a simulation. These values are referred to as the structure identification number $\mathcal{S}$ and the liquid identification number $\mathcal{L}$. Every liquid structure in the simulation domain is tagged with each of these values. The process in which $\mathcal{S}$ and $\mathcal{L}$ are assigned, transported, and reassigned is the basis for identifying liquid splitting and merging events within a simulation. 

\subsubsection{Identification Numbers} \label{id_num}
Liquid identification numbers ($\mathcal{L}$) are transported with the liquid and provide a time history of where liquid moves.  Therefore, the $\mathcal{L}$ values are not reassigned at every timestep. They are persistent through time and are only reassigned after a split event creates a new liquid structure or a coalescence event destroys a liquid structure. $\mathcal{L}$ values are transported with the liquid they occupy, so $\mathcal{L}$ transport must be tied into the simulation's liquid transport scheme. In our implementation, $\mathcal{L}$ fluxes are calculated in the same manner as the semi-Lagrangian liquid fluxes in NGA's geometric VOF scheme~\cite{Owkes2014,Owkes2017}. A list ($\mathcal{L}_\mathrm{Flux}$) of all the $\mathcal{L}$ that flux into a cell during a timestep is constructed. This is done by first constructing lists at each cell face with signed $\mathcal{L}$ values indicating the direction. After lists associated with each cell face are computed, they are compiled, to create a list of all $\mathcal{L}$'s within a cell. These lists are looped through in a later section of the code to identify coalescence. Fig.~\ref{fig:LID_flux} displays a visual representation of this process. See \ref{appendixa} for the $\mathcal{L}$ transport and coalescence identification algorithm. 

Structure identification numbers ($\mathcal{S}$) are assigned at every timestep following liquid transport. The values of $\mathcal{S}$ are not consequential, provided that every structure is assigned a unique value. Independent structures are identified and tagged using a band-growth algorithm first described by Herrmann~\cite{Herrmann2010}. In the present application, the algorithm operates as follows. To begin, $\mathcal{S}$ is assigned on the entire domain to be zero. Then, the domain is looped through to find the first untagged liquid node, meaning the liquid volume fraction is greater than zero and $\mathcal{S}$ = 0. This point is then tagged with an $\mathcal{S}$ and the cells around that node are looped through until the entire continuous structure is identified and assigned the $\mathcal{S}$ value. A logical statement determines whether an adjacent cell will be assigned the same $\mathcal{S}$ value as the present cell. The statement requires that: 1) there must be liquid in that cell, 2) it must be untagged ($\mathcal{S} = 0$), and  3) the $\mathcal{L}$ of the adjacent cell must equal that of the current cell. The third statement ensures that liquid structures do not re-merge immediately following breakup as structures advance and naturally inhabit adjacent cells. See Section~\ref{fictitious_e} for a more detailed description of these criteria. Herrmann's work also described the parallelization of this process, which is utilized in the extraction tool. This requires special consideration be given to communicating $\mathcal{S}$ between processors. Many other methods exist to assign the structure identification number. This type of tagging problem is known as Connected Component Labeling, e.g.~\cite{Hendrickson2020,Hoshen1976}.  

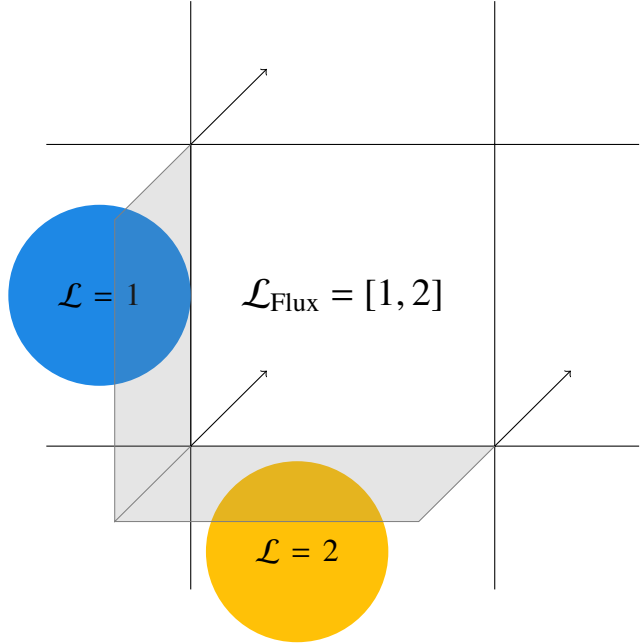
\begin{figure}[htbp]
 \centering
 \begin{tikzpicture}
    \draw[step=4cm,thin] (-1.9,-1.9) grid (5.9,5.9);
    \draw (0,0) -- (0,4);
    \fill [fill={myblue}] (-1.2,2) circle (1.2);
    \fill [fill={myellow}] (1.4,-1.4) circle (1.2);
    \node [text width=1.5cm,align=center] at (-1.2,2) {\large$\mathcal{L}=1$};
    \node [text width=1.5cm,align=center] at (1.4,-1.4) {\large$\mathcal{L}=2$};
    \filldraw[gray, opacity = 0.2] (0,4) -- (-1,3) -- (-1,-1) -- (0,0);
    \draw [help lines] (0,4) -- (-1,3) -- (-1,-1) -- (0,0);
    \draw [help lines] (-1,-1) -- (3,-1) -- (4,0);
    \filldraw[gray, opacity=0.2] (-1,-1) -- (3,-1) -- (4,0) -- (0,0);
    \node [text width=5cm,align=center] at (2,2) {\Large $\displaystyle \mathcal{L}_\mathrm{Flux}\!\!\!=\!\!\![1,2]$};
    \draw[->] (4,0) -- (5,1);
    \draw[->] (0,0) -- (1,1);
    \draw[->] (0,4) -- (1,5);
  \end{tikzpicture}
 \caption{An example of two droplets fluxing into the same cell. Gray regions are semi-lagrangian flux regions~\cite{Owkes2014}, which carry $\mathcal{L}$ into the center cell due to the velocity shown with vectors. The $\mathcal{L}$'s of the two droplets are stored in the list $\mathcal{L}_\mathrm{Flux}$.}
 \label{fig:LID_flux}
\end{figure}

\subsubsection{Breakup Identification}
\begin{figure}[htbp]
 \centering
 \includegraphics[width=1\textwidth]{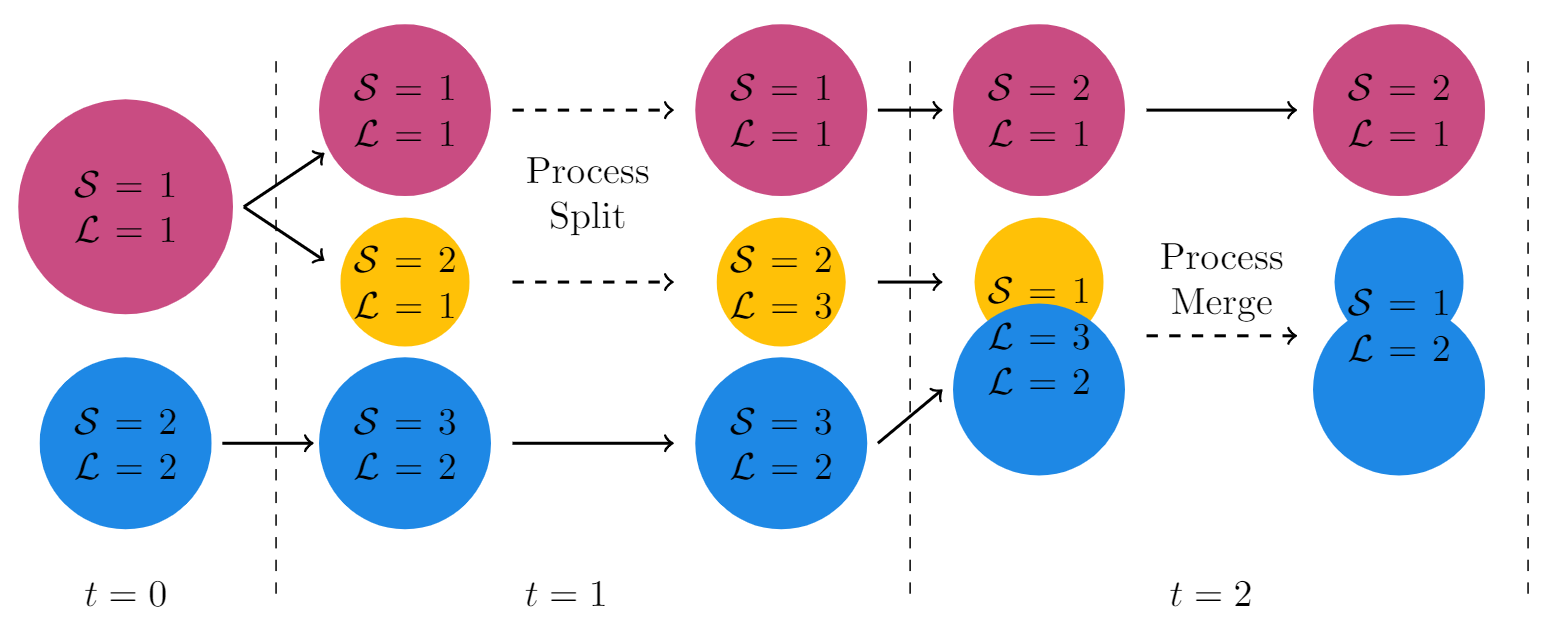}
 \caption{An example of two droplets undergoing breakup and coalescence. Two droplets start on the left,  the top droplet breaks into two droplets.  The two droplets are identified and the split event is processed.  The bottom two droplets coalesce, the merge is identified and the merge event is processed. Note that the $\mathcal{S}$ can vary between timesteps, but the $\mathcal{L}$ is persistent until a split or merge causes the value to change.}
 \label{fig:splitmerge}
\end{figure}

Liquid structure splitting or breakup is identified when two different structures (two unique $\mathcal{S}$'s) have the same $\mathcal{L}$. This $\mathcal{L}$ is persistent through time, indicating that these two droplets were part of the same structure at the previous timestep. The logical statement, described in Section~\ref{id_num}, loops over adjacent cells to connect liquid structures. Two structures become independent of each other when they are not in adjacent cells i.e., there is a full cell separating them. This is when the $\mathcal{S}$ assignment portion of the code assigns two structures with unique $\mathcal{S}$'s. The two structures have the same $\mathcal{L}$, indicating that a breakup occurred. When a breakup is identified, data are extracted from the event, and a new $\mathcal{L}$ is assigned to the smaller structure(s) created by the breakup event. See $t=0$ and $t=1$ in Fig~\ref{fig:splitmerge} for a visual representation of this process.  

\subsubsection{Coalescence Identification}\label{coalescence}
Coalescence of liquid structures is identified using the $\mathcal{L}_{flux}$ lists mentioned in Section~\ref{id_num}. More than one unique $\mathcal{L}$ in a cell indicates that two structures are merging together. When coalescence is identified, data are extracted from the event, and the new merged structure takes the smaller $\mathcal{L}$ of the two parent structures.

\subsubsection{Addressing Fictitious Events} \label{fictitious_e}
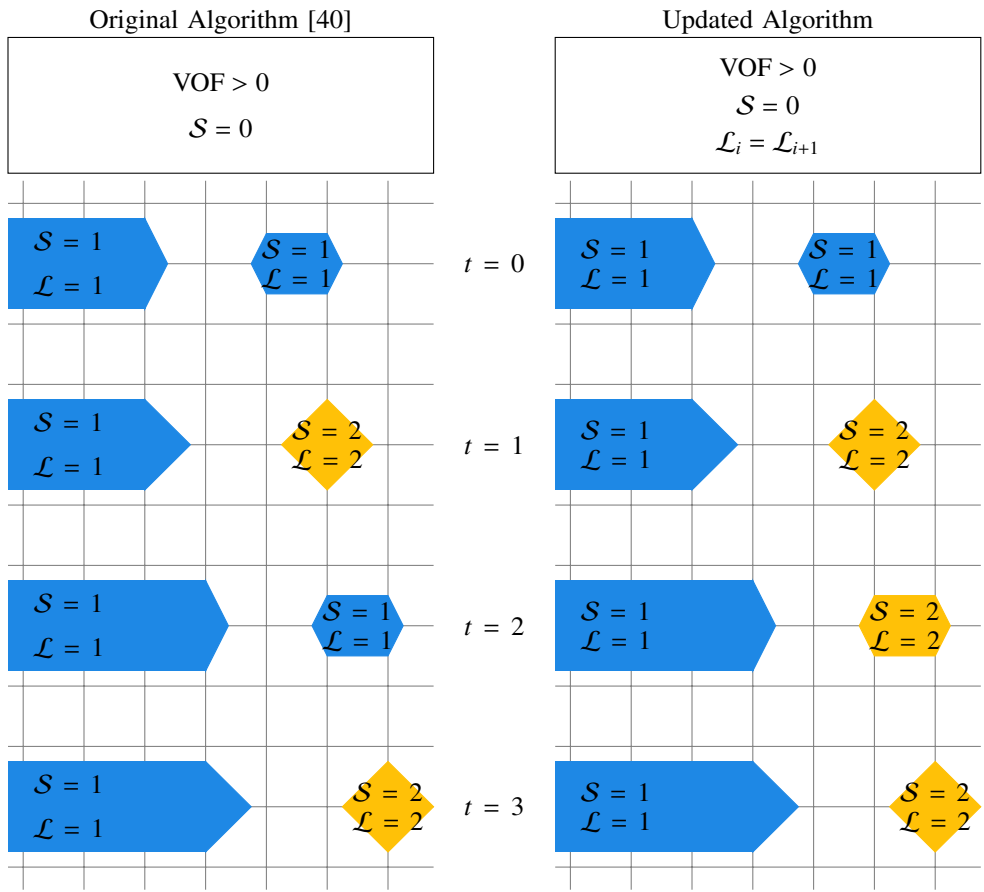
\begin{figure}
  \centering
    \begin{tikzpicture}
      \node [text width=1.5cm,align=center] at (3.8,0) {$t = 3$};
      \node [text width=1.5cm,align=center] at (3.8,2.4) {$t = 2$};
      \node [text width=1.5cm,align=center] at (3.8,4.8) {$t = 1$};
      \node [text width=1.5cm,align=center] at (3.8,7.2) {$t = 0$};
      \node [text width=4cm,align=center] at (0.2,10.4) {Original Algorithm~\cite{RubelOwkes2019}};
      \node [text width=4cm,align=center] at (7.4,10.4) {Updated Algorithm};
      \draw (-2.6,8.4) -- (3,8.4) -- (3,10.2) -- (-2.6,10.2) -- cycle;
      \node at (0.2,9.6) {VOF $> 0$};
      \node at (0.2,9) {$\mathcal{S} = 0$};
      \draw (4.6,8.4) -- (10.2,8.4) -- (10.2,10.2) -- (4.6,10.2) -- cycle;
      \node at (7.4,9.8) {VOF $> 0$};
      \node at (7.4,9.3) {$\mathcal{S} = 0$};
      \node at (7.4,8.8) {$\mathcal{L}_i$ = $\mathcal{L}_{i+1}$};
      \draw [step=0.8cm,gray,thin] (-2.6,-1.1) grid (3,8.3);
      \fill [{myblue}] (-2.6,-0.6) rectangle (0,0.6);
      \filldraw [{myblue}] (0,-0.6) -- (0.6,0) -- (0,0.6) -- cycle;
      \node [text width=1.5cm,align=center] at (-1.8,0.3) {$\mathcal{S}=1$};
      \node [text width=1.5cm,align=center] at (-1.8,-0.3) {$\mathcal{L}=1$};
      \filldraw [{myellow}] (1.8,0) -- (2.4,-0.6) -- (3,0) -- (2.4,0.6) -- cycle;
      \node [text width=1.5cm,align=center] at (2.4,0.2) {$\mathcal{S}=2$};
      \node [text width=1.5cm,align=center] at (2.4,-0.2) {$\mathcal{L}=2$};
      Row    
      \fill [{myblue}] (-2.6,1.8) rectangle (0,3);
      \filldraw [{myblue}] (0,1.8) -- (0.3,2.4) -- (0,3) -- cycle;
      \node [text width=1.5cm,align=center] at (-1.8,2.7) {$\mathcal{S}=1$};
      \node [text width=1.5cm,align=center] at (-1.8,2.1) {$\mathcal{L}=1$};
      \filldraw [{myblue}] (1.4,2.4) -- (1.6,2) -- (2.4,2) -- (2.6,2.4) -- (2.4,2.8) -- (1.6,2.8) -- cycle;
      \node [text width=1.5cm,align=center] at (2,2.6) {$\mathcal{S}=1$};
      \node [text width=1.5cm,align=center] at (2,2.2) {$\mathcal{L}=1$};
      \fill [{myblue}] (-2.6,4.2) rectangle (-0.8,5.4);
      \filldraw [{myblue}] (-0.8,4.2) -- (-0.2,4.8) -- (-0.8,5.4) -- cycle;
      \node [text width=1.5cm,align=center] at (-1.8,5.1) {$\mathcal{S}=1$};
      \node [text width=1.5cm,align=center] at (-1.8,4.5) {$\mathcal{L}=1$};
      \filldraw [{myellow}] (1,4.8) -- (1.6,4.2) -- (2.2,4.8) -- (1.6,5.4) -- cycle;
      \node [text width=1.5cm,align=center] at (1.6,5.0) {$\mathcal{S}=2$};
      \node [text width=1.5cm,align=center] at (1.6,4.6) {$\mathcal{L}=2$};
      \fill [{myblue}] (-2.6,6.6) rectangle (-0.8,7.8);
      \filldraw [{myblue}] (-0.8,6.6) -- (-0.5,7.2) -- (-0.8,7.8) -- cycle;
      \node [text width=1.5cm,align=center] at (-1.8,7.5) {$\mathcal{S}=1$};
      \node [text width=1.5cm,align=center] at (-1.8,6.9) {$\mathcal{L}=1$};
      \filldraw [{myblue}] (0.6,7.2) -- (0.8,6.8) -- (1.6,6.8) -- (1.8,7.2) -- (1.6,7.6) -- (0.8,7.6) -- cycle;
      \node [text width=1.5cm,align=center] at (1.2,7.4) {$\mathcal{S}=1$};
      \node [text width=1.5cm,align=center] at (1.2,7.0) {$\mathcal{L}=1$};
     
      \draw [step=0.8cm,gray,thin] (4.6,-1.1) grid (10.2,8.3);
      \fill [{myblue}] (4.6,-0.6) rectangle (7.2,0.6);
      \filldraw [{myblue}] (7.2,-0.6) -- (7.8,0) -- (7.2,0.6) -- cycle;
      \node [text width=1.5cm,align=center] at (5.4,0.2) {$\mathcal{S}=1$};
      \node [text width=1.5cm,align=center] at (5.4,-0.2) {$\mathcal{L}=1$};
      \filldraw [{myellow}] (9,0) -- (9.6,-0.6) -- (10.2,0) -- (9.6,0.6) -- cycle;
      \node [text width=1.5cm,align=center] at (9.6,0.2) {$\mathcal{S}=2$};
      \node [text width=1.5cm,align=center] at (9.6,-0.2) {$\mathcal{L}=2$};
      \fill [{myblue}] (4.6,1.8) rectangle (7.2,3);
      \filldraw [{myblue}] (7.2,1.8) -- (7.5,2.4) -- (7.2,3) -- cycle;
      \node [text width=1.5cm,align=center] at (5.4,2.6) {$\mathcal{S}=1$};
      \node [text width=1.5cm,align=center] at (5.4,2.2) {$\mathcal{L}=1$};
      \filldraw [{myellow}] (8.6,2.4) -- (8.8,2) -- (9.6,2) -- (9.8,2.4) -- (9.6,2.8) -- (8.8,2.8) -- cycle;
      \node [text width=1.5cm,align=center] at (9.2,2.6) {$\mathcal{S}=2$};
      \node [text width=1.5cm,align=center] at (9.2,2.2) {$\mathcal{L}=2$};
      \fill [{myblue}] (4.6,4.2) rectangle (6.4,5.4);
      \filldraw [{myblue}] (6.4,4.2) -- (7,4.8) -- (6.4,5.4) -- cycle;
      \node [text width=1.5cm,align=center] at (5.4,5.0) {$\mathcal{S}=1$};
      \node [text width=1.5cm,align=center] at (5.4,4.6) {$\mathcal{L}=1$};
      \filldraw [{myellow}] (8.2,4.8) -- (8.8,4.2) -- (9.4,4.8) -- (8.8,5.4) -- cycle;
      \node [text width=1.5cm,align=center] at (8.8,5.0) {$\mathcal{S}=2$};
      \node [text width=1.5cm,align=center] at (8.8,4.6) {$\mathcal{L}=2$};
      \fill [{myblue}] (4.6,6.6) rectangle (6.4,7.8);
      \filldraw [{myblue}] (6.4,6.6) -- (6.7,7.2) -- (6.4,7.8) -- cycle;
      \node [text width=1.5cm,align=center] at (5.4,7.4) {$\mathcal{S}=1$};
      \node [text width=1.5cm,align=center] at (5.4,7.0) {$\mathcal{L}=1$};
      \filldraw [{myblue}] (7.8,7.2) -- (8,6.8) -- (8.8, 6.8) -- (9,7.2) -- (8.8,7.6) -- (8,7.6) -- cycle;
      \node [text width=1.5cm,align=center] at (8.4,7.4) {$\mathcal{S}=1$};
      \node [text width=1.5cm,align=center] at (8.4,7.0) {$\mathcal{L}=1$};
    \end{tikzpicture}
    \caption{The diagram on the left shows the issue with fictitious merging and splitting events in the original algorithm~\cite{RubelOwkes2019}. The diagram to the right shows how the split and merge identifier works in the present work. Notice that at $t = 2$ in the updated code the structures do not merge back together. This is because the criterion for coalescence was changed to require the structures to exist within the same cell, not adjacent cells.}
    \label{fig:updates}
\end{figure}

An issue from the previous work, described in Rubel and Owkes~\cite{RubelOwkes2019} was the occurrence of numerous fictitious merge and split events. These occurred when the tool identified a structure that broke into two structures and then the two structures coalesced. The algorithm identified this process and repeated many times, when in fact the liquid structures never re-merged following breakup. 
The error stemmed from the original $\mathcal{S}$ assignment algorithm. Droplets in adjacent cells were automatically treated as the same structure, which resulted in recently split liquid structures being incorporated back into their parent droplet when they were in neighboring cells. To avoid an accumulation of non-physical results, the authors opted to run the split identification portion of the tool only every 10 timesteps. The present work changed the parameters that define a coalescence event to prevent droplets from merging together immediately after breakup. This allows the tool to run every timestep, bolstering it's accuracy and usefulness. See Fig.~\ref{fig:updates} for a visualization of this issue and the result following the updates in this work.

As stated above, in Section~\ref{coalescence}, coalescence now occurs when two structures flux into the same cell. This methodology requires that special care be given to the $\mathcal{S}$ assignment algorithm to ensure that separate structures in adjacent cells are not combined. To address this, the $\mathcal{L}$ of droplets was incorporated into the band-growth algorithm. Fig.~\ref{fig:updates} illustrates this updated $\mathcal{S}$ assignment process. At $t = 0$ a droplet is breaking up so that there is liquid in adjacent cells. Since a split has not been identified yet, the $\mathcal{L}$ of both cells is the same, resulting in a continuous structure. Then, at $t = 1$ the structure on the right moves, leaving a full cell between it and the other structure. The cells adjacent to the blue structure are void of liquid, so the $\mathcal{S}$ value is only assigned to that structure. Since the droplet on the right becomes independent, it is assigned a new $\mathcal{S}$, a split is processed and it is also assigned a new $\mathcal{L}$. Finally, at $t = 2$, previously, the two droplets would have merged back together as they occupy neighboring cells.  But the structures remain separate with the new algorithm, because the $\mathcal{L}$ of the adjacent droplet does not equal the $\mathcal{L}$ of the blue droplet, they remain separate structures and will not coalesce unless they enter the same cell.
\label{sec:fict_events}

\subsection{Data Extraction}
Following identification of breakup or coalescence in the simulation, data are extracted from that location and time. This is intended to provide access to the local conditions of these events. The original work extracted data from the timestep immediately following breakup. We expanded the temporal sampling range, allowing the tool to extract data from the timestep immediately preceding breakup in addition to the timestep following breakup. This improvement is intended to provide insight into the conditions which lead to breakup and to the types of structures which the conditions produce. To highlight some potential uses of the tool, the present work extracted the locations of the events, the gas and liquid velocities, droplet volume, and droplet shape information. 

For the purposes of this study, liquid and gas velocities are calculated as volume averaged velocities,
\begin{align}
    \bm{U}_\mathrm{liquid} &= \frac
    {\sum_{i=1}^{N_{s}}{\bm{u}_{i} \mathcal{V}_{\mathrm{cell},i} ~\alpha_i}}
    {\sum_{i=1}^{N_{s}}{\mathcal{V}_{\mathrm{cell},i} ~\alpha_i}}\\
    \bm{U}_\mathrm{gas} &= \frac
    {\sum_{i=1}^{N_{s}}{\bm{u}_{i} \mathcal{V}_{\mathrm{cell},i} ~(1-\alpha_i)}}
    {\sum_{i=1}^{N_{s}}{\mathcal{V}_{\mathrm{cell},i} ~(1-\alpha_i)}}
\end{align}
respectively. These values are calculated by considering all cells that contain a liquid structure and all cells adjacent to those ($N_s$ total cells). In the expressions, $\bm{u}_{i}$ is the velocity vector, $\mathcal{V}_{\mathrm{cell},i}$ is the cell volume, and $\alpha_i$ is the liquid volume fraction in the ${i}^\mathrm{th}$ cell. This method for calculating the velocities is preliminary and does not fully capture the local flow field dynamics. Future work focused on extracting topological data from the flow field will work to better quantify these values. 

\subsection{Graphical Database}
The primary goal of the tool is to make data from atomization simulations more accessible. Special consideration was given to ensuring that extracted data was stored in an efficient and easily queriable format. We opted to store the atomization data in a Neo4j graphical database. Graph databases are commonly used in the corporate sector for companies to create connections between users and products through paths, which can reveal patterns in group dynamics and buying trends. The same principles can be applied to atomization simulations through the construction of paths which connect droplets through breakup and coalescence events. Storing atomization data in a graph database format allows researchers to analyze the evolution of liquid as it moves from the liquid core to small droplets in an atomizing flow. In this work, the Neo4j database is used to store the atomization data. Data from the simulation in this work is initially written to a CSV file. Then, rather than simply storing the data as a disconnected list of events, the CSV file is uploaded to Neo4j. The graph format presents an easy way to connect related events and develop a droplet ancestry. Additionally, storing data in graph databases provide a unique way to visualize atomization data and a novel method for studying the atomization process.
\begin{figure}[htbp]
 \centering
 \includegraphics[width=1\textwidth]{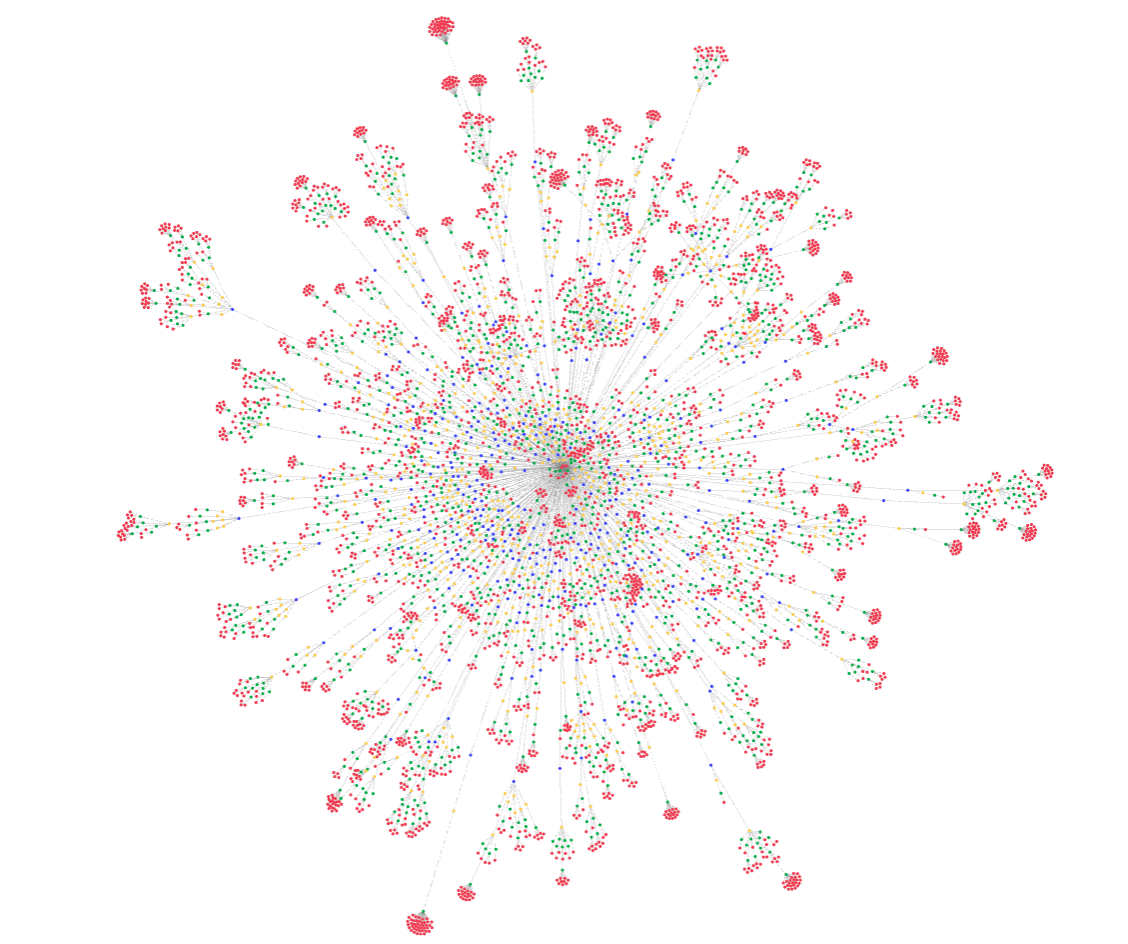}
 \caption{Example of data points in Neo4j graphical database. Red nodes represent some droplets which broke up 4 times throughout the simulation. Other colors represent intermediate breakups (blue = primary, yellow = secondary, green = tertiary). Lines connect related droplets, i.e., droplets which split from each other. The liquid core is present at the center of the image. Each node contains relevant statistics from the breakup event.} 
 \label{fig:Neo4j}
\end{figure}

Fig.~\ref{fig:Neo4j} displays how data are stored in a Neo4j database. This image represents some of the data from the simulation described in Section 3. The nodes are droplets produced from breakup events. The breakup events are represented by the lines connecting the nodes. The node colors represent the number of times they have broken up. The liquid core is in the center of the image and the red droplets represent the sixth breakup events. This image is made up of all the droplets which broke up 6 times during the first 2.3 $\mu$s of the simulation, and all the intermediate droplets between the coherent liquid core and these sixth breakup droplets. Within each node is stored the extracted data from the point and time in the simulation in which that droplet broke up and became independent. So, this database is saturated with data, but organized in a logical and accessible format.

The graph database can be efficiently edited, reorganized, and parsed with the Cypher Query Language. It is syntactically similar to SQL, but designed to specifically query nodes and relationships and the paths formed with these components. As mentioned above, data is imported to Neo4j via a CSV file, organized in such a way that each row is an independent liquid breakup or coalescence event. These rows are imported into Neo4j as droplet nodes and breakup and coalescence relationships are created. See \ref{appendixb} for a detailed description of the process to import data into Neo4j. Following data import, Cypher can be used to further organize the data and/or query the data to analyze the atomization process. Below are some examples of the capabilities of Cypher in the present application. 

\begin{lstlisting}
// Rename the liquid core
MATCH(n:droplet)     **// Find nodes with the "droplet" label**
WHERE n.Event='None' **// Node with Event = none is the liquid core**  
CALL apoc.refactor.rename.label("droplet","core",n)  **// Rename node**
YIELD committedOperations
RETURN committedOperations

// Identify and rename primary droplets
MATCH (n:droplet)
WHERE n.OldLID = 1   **// Find droplets which broke off of the liquid core**
WITH collect(n) as p **// Compile a list of nodes matching criteria**
CALL apoc.refactor.rename.label("droplet","primary",p)
YIELD committedOperations
RETURN committedOperations

// Rename secondary,tertiary, etc. (repeat until no "droplet" nodes remain)
MATCH (n:droplet),(d:primary)
WHERE n.OldLID = d.NewLID **// Find "droplet" nodes that split from "primary"**
WITH collect(n) as p
CALL apoc.refactor.rename.label("droplet","secondary",p)
YIELD committedOperations
RETURN committedOperations
\end{lstlisting}

This portion of code first identifies the core by finding the only node which has an event not equal to either breakup or coalescence. Then, it renames this droplet "core". Note: the "Awesome Procedures on Cypher" (APOC) library must be enabled to access the renaming features. Next, primary droplets are identified by finding all the droplet nodes which previously had a $\mathcal{L}$ (LID) equal to one (the liquid core has $\mathcal{L} = 1$). These droplets are renamed "primary". After primary droplets are identified, secondary droplets can be identified using a similar process. The algorithm loops through all droplets and looks for nodes which previously had $\mathcal{L}$ equal to the $\mathcal{L}$ of a primary droplet. This process can then be repeated for each subsequent breakup event until no more "droplet" nodes remain. This is very useful in order to analyze the evolution of droplets as they break up further. This will be analyzed in more detail in the Results section. 

\begin{lstlisting}
// Collect all breakup paths of droplets which break up six times
MATCH (n:sixth), (c:core), p = shortestPath((c)-[:Split*]->(n))
RETURN [d in nodes(p)| d.Volume]
\end{lstlisting}

The "shortestPath" function is used to extract the paths between specific nodes. In the above example, the shortest path from the liquid core to droplets that broke up six times is queried. Then, the volume of each droplet in that path is returned. This feature provides a simple method to extract useful relational data from the breakup process.

In addition to the Neo4j graphical database system and the Cypher Query Language, we utilized Python for analysis of results. Both programs have benefits and drawbacks, which led to the utilization of both. Neo4j allowed us to build paths very easily between droplets which break up from each other or coalesce together. From this information, we can easily output a CSV which displays how statistics evolve with breakup events or coalescence events. Following the output of these CSV files, it is simple to analyze the data using Python's data science libraries. Additionally, multiple options exist to query or create a graph databases through Python. These include the official Neo4j driver for Python and the py2neo Python library. These prove to be very powerful tools, because they allow users to combine python loops, if statements, and data science libraries with the unique organizational system of Neo4j. The py2neo library was used in the present work to build the time-series plots in section~\ref{sec:time}. See appendix~\ref{appendixc} for an example python script querying a Neo4j database.

\section{Application of Extraction Tool on Diesel Jet}

\subsection{Simulation Setup}
To test the utility of the updated tool we ran a simulation inspired by a diesel injector.  The simulation is the same as that described in Rubel and Owkes~\cite{RubelOwkes2019} and consists of a turbulent liquid injection into a quiescent air.  The turbulence was computed from a preliminary turbulent pipe-flow simulation and the velocity field was stored and used as the inlet boundary condition for the liquid jet.  Table~\ref{Table:3Djetnum} provides the parameters of the simulated jet. The simulation was run on 160 processors on the Hyalite High-Performance Computing Cluster at Montana State University. The dimensions of the computational mesh are $\mathcal{N}_x\times\mathcal{N}_y\times\mathcal{N}_z=1024\times128\times128$. Note that the resolution of this simulation is not fine enough to accurately capture small scale interface features or the smallest scales of turbulence. However, this simulation is sufficient to demonstrate the efficacy of the proposed tool. Future work will focus on applying the tool to a higher resolution simulation to more accurately extract information on physical phenomena.

\begin{table}[htbp]
\begin{center}
\begin{tabular}{ |c|c|c| } 
 \hline
 Number & Definition & Value \\
 \hline
 Bulk Reynolds number & $\rho_{l}U_{\mathrm{jet}}D_{\mathrm{jet}}/\mu_{l}$ & 25,000 \\ 
 Bulk Weber number & $\rho_{l}U_{\mathrm{jet}}^2D_{\mathrm{jet}}/\sigma_{l}$ & 10,000 \\ 
 Density ratio & $\rho_{l}/\rho_{g}$ & 40 \\
 Viscosity ratio & $\mu_{l}/\mu_{g}$ & 1.67\\
 Domain length & $L_{x}/D_{\mathrm{jet}}$ & 60 \\
 Domain widths & $L_{y,z}/D_{\mathrm{jet}}$ & 7.5 \\
 Cells across diameter & $D_{\mathrm{jet}}/\Delta x$ & 17.06 \\
 CFL number & $|u|_{\mathrm{max}}\Delta t/\Delta x$ & 0.4 \\
 \hline
\end{tabular}
\end{center}
\caption{Non-dimensional parameters used in the diesel jet simulation.}
\label{Table:3Djetnum}
\end{table}

\begin{figure}[htbp]
 \centering
 \includegraphics[width=1\textwidth]{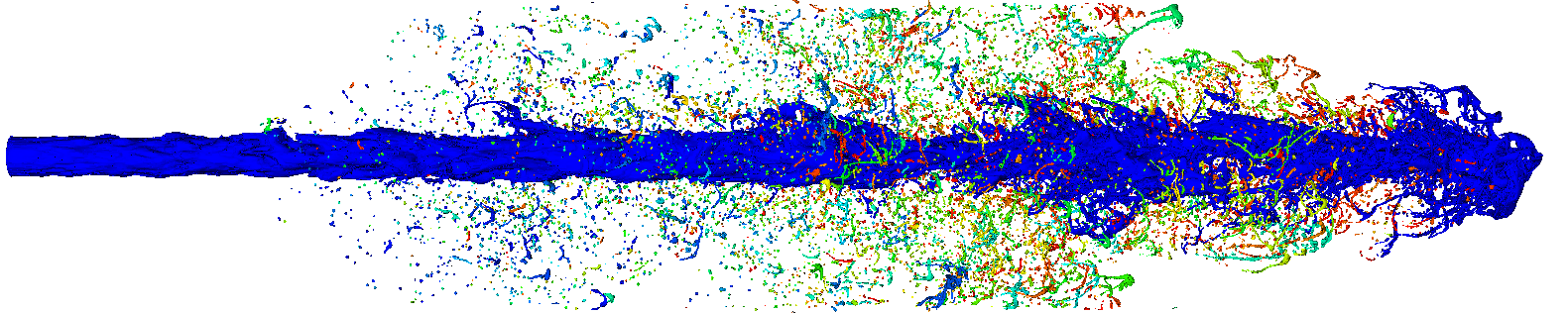}
 \caption{Rendering of the liquid jet run in simulation. $\mathcal{L}$ values at final timestep represented by colors. Rendered using VisIt~\cite{HPV:VisIt}.} 
 \label{fig:jet}
\end{figure}

\subsection{Addressing Fictitious Events Confirmation}
Section~\ref{sec:fict_events} discusses an update in the coalescence and breakup identification algorithm, which is intended to prevent the occurrence of fictitious events. In the original work~\cite{RubelOwkes2019}, the fictitious events coincided with every breakup within the simulation, which forced the authors to institute a delay in the breakup identification portion of the code. This meant they only identified breakup events every 10 timesteps. In the present work, breakup identification is performed on every timestep. It is reasonable in complex flow systems that droplets may breakup and then re-merge with their parent droplet, so the algorithm is not intended to prevent this phenomena from occurring altogether. 

To confirm that the updated method is producing reasonable results, we utilized Neo4j to identify all the droplets which split and re-merged with their parent droplet. Then, the difference in time between the coalescence event and the initial breakup event was calculated. 
17.93\% of all droplets re-merged with their parent droplet, with the majority of those being primary droplets coalescing with the liquid core. This is a well documented mechanism~\cite{Shinjo2010} and expected in a round jet injected into quiescent gas. Analyzing only secondary droplets, we found that 9.31\% of secondary droplets re-merged with their parent droplets. The fictitious events were documented to occur within one or two timesteps of breakup, but in the present simulation, the average time between coalescence and breakup is $61.7$ $\mu$s which translates to about 300 timesteps. Given these statistics, we conclude that the updates made to the tool in the present work remedied the fictitious breakup and coalescence issue addressed in the original work and the identified events in the present work are consistent with physical events in the spray.   

\subsection{Secondary Atomization Analysis}
\begin{figure}[htbp]
 \centering
 \includegraphics[width=1\textwidth]{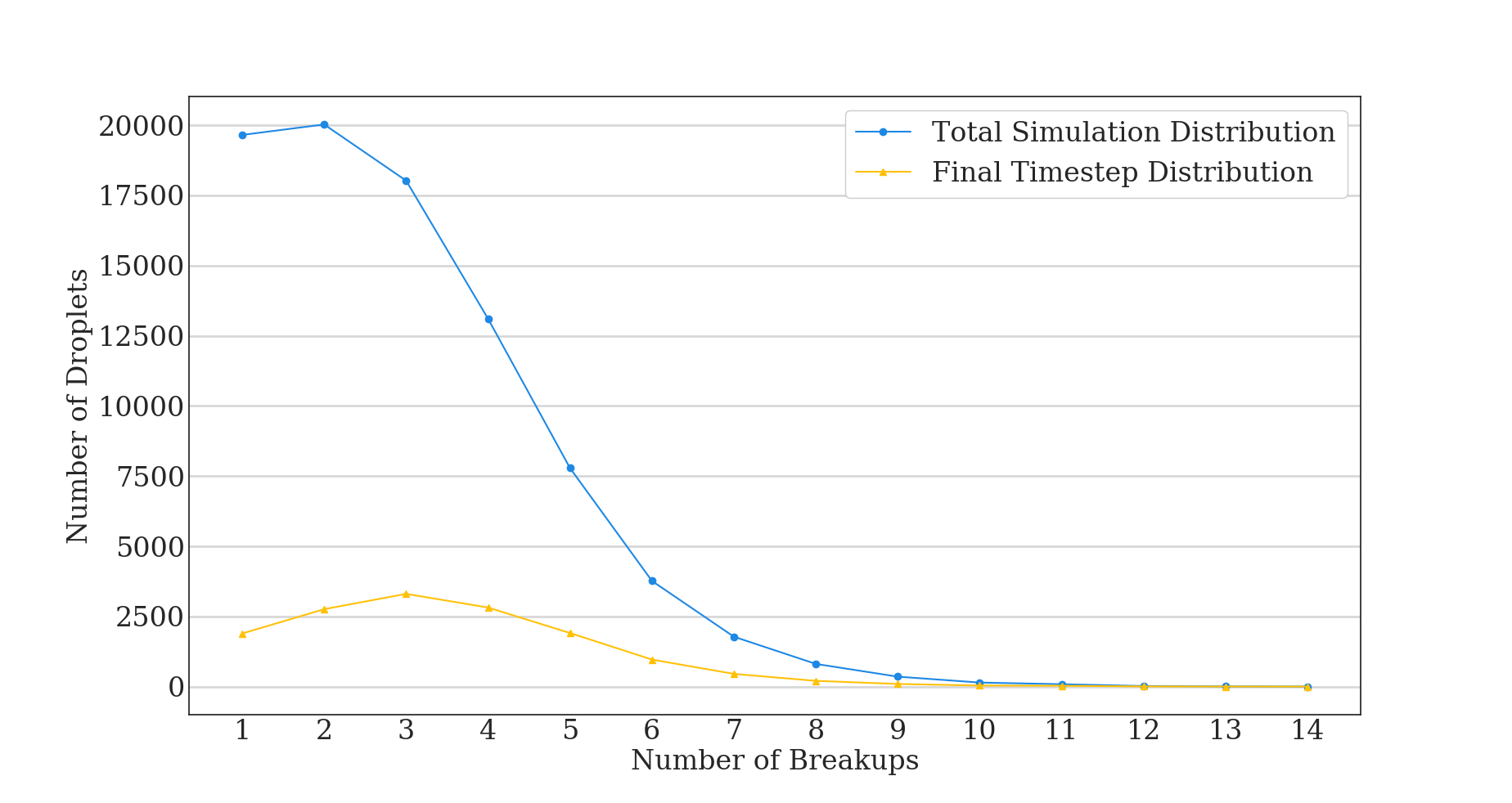}
 \caption{The total number of droplets associated with each breakup stage from all breakup events within the diesel jet simulation.} 
 \label{fig:sec_hist}
\end{figure}

Common descriptions of breakup regimes within an atomizing system involve discussion of primary atomization, i.e. droplets splitting from the liquid core, and secondary atomization, i.e., all further breakup. As seen in Fig.~\ref{fig:sec_hist}, many droplets are created after multiple breakup events.  Further analysis shows that 53.6\% of the total number of droplets formed in the test diesel jet broke up three or more times, with 17.0\% percent breaking up at least five times. At later times in the simulation, this percentage becomes more pronounced. The yellow curve with triangle markers in Fig.~\ref{fig:sec_hist} is the distribution of droplets at the final timestep. 67.7\% of drops present at this timestep broke up 3 or more times. Thus, a majority of the breakup in the simulation occurred after secondary breakup. This indicates that the final spray field is heavily dependent on mechanisms within these later atomization regimes, which are all generally grouped in secondary atomization.

\subsection{Coalescence Analysis}
A trend extracted from the test case simulation is the prevalence of coalescence events. This matches findings in Prakash et al.~\cite{Prakash2019}, who identified a significant number of coalescence events within their simulation. The present simulation produced 76,608 breakup events and 57,908 coalescence events. Table~\ref{Table:merge} displays the average number of coalescence events for a given number of breakup events. These data were calculated using a path finding algorithm in Neo4j. The algorithm finds the shortest path of breakup events from droplets at the given breakup stage back to the liquid core and counts the number of coalescence events along that path. The average number of coalescence events shown in the table is the average amongst all droplets that have broken up the associated number of times.  The data in Table~\ref{Table:merge} indicates that the proportion of coalescence events increases as the jet evolves and more breakup occurs. This makes intuitive sense, because as the jet evolves, a high droplet density cloud develops, which leads to more droplet collisions.

\begin{table}[htbp]
\begin{center}
\begin{tabular}{ |c|c|c| } 
 \hline
 \thead{Breakup Stage} & \thead{Average Number of\\Coalescence Events} & \thead{Ratio of Coalescence\\to Breakup}\\
 \hline
 Secondary & 1.077 & 0.5385\\ 
 \hline
 Third & 1.899 & 0.6330\\
 \hline
 Fourth & 2.898 & 0.7245\\
 \hline
 Fifth & 3.974 & 0.7948\\
 \hline
 Sixth & 5.241 & 0.8735\\
 \hline
 Seventh & 6.823 & 0.9747\\
 \hline
 Eighth & 8.276 & 1.035\\
 \hline
\end{tabular}
\end{center}
\caption{The average number of coalescence events between primary breakup and the breakup stage is listed in the second column. The third column displays the ratio of coalescence events to breakup events along breakup paths.}
\label{Table:merge}
\end{table}

\subsection{Local Flow Field Statistics}
\begin{figure}[!htbp]
 \centering
 \includegraphics[width=1\textwidth]{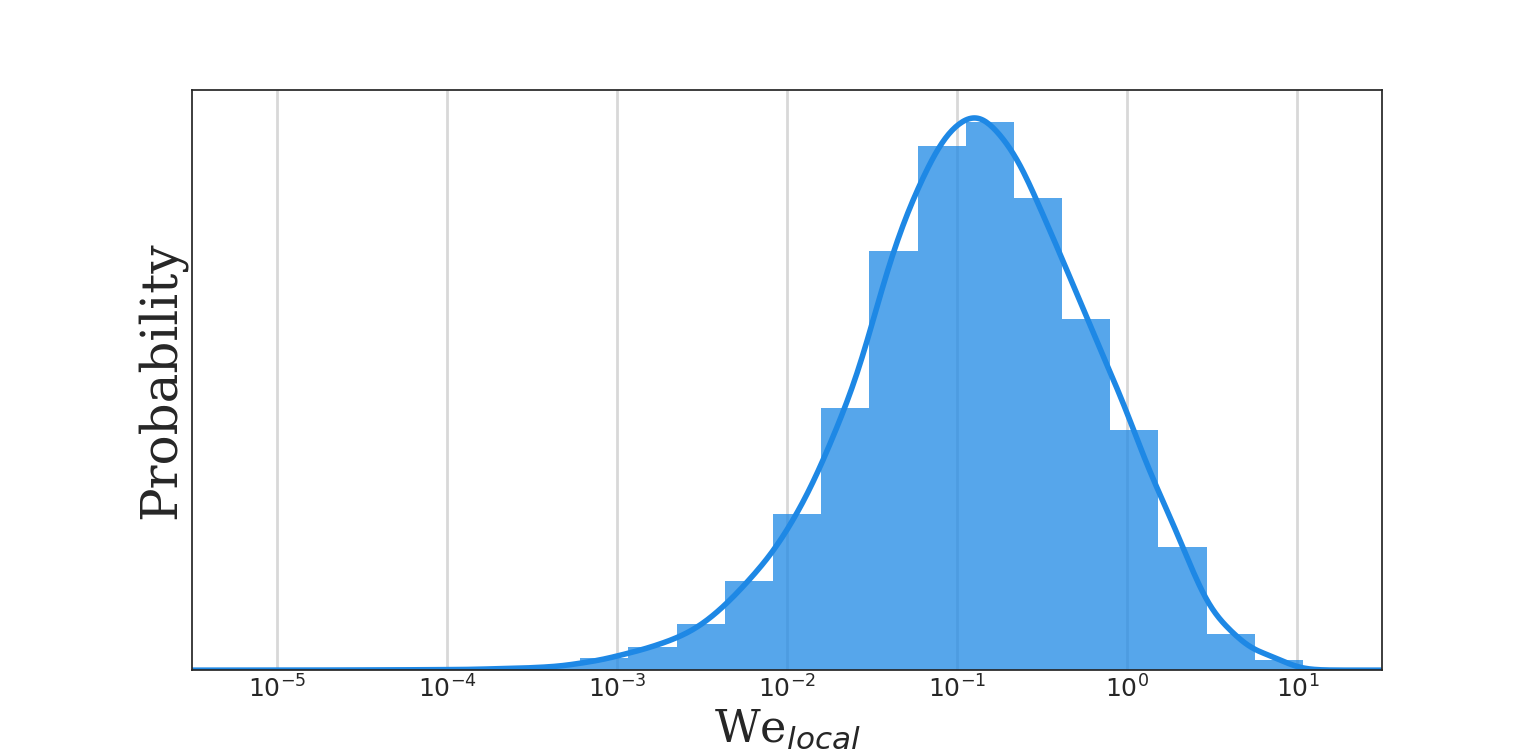}
 \caption{A probability density function of the local Weber Number associated with every breakup in the simulation.} 
 \label{fig:weber_pdf}
\end{figure}

Another use for this tool is to better understand how the local flow field is affecting the atomization process. This is information that is exceedingly difficult, if not impossible, to obtain through experimental methods. Moreover, gathering a sufficient sample of statistics on these local flow mechanisms has remained nonviable for those conducting numerical simulations. This work extracted preliminary data on the local flow field surrounding breakup events and calculated the resultant local Weber number, defined as $\mathrm{We}_\mathrm{local}={\rho_g U_s^2 L}/{\sigma}$, where $\rho_g$ is the gas velocity, $U_s$ is the slip velocity (defined as $U_s = |U_\mathrm{liquid}-U_\mathrm{gas}|$), $L$ is the characteristic length (in this case, it is the equivalent spherical diameter of the droplet prior to break up), and $\sigma$ is the surface tension coefficient. 
A probability density function (PDF) of the logarithm of $\mathrm{We}_\mathrm{local}$ number is developed from every breakup within the simulation (Fig.~\ref{fig:weber_pdf}). The values are centered roughly about $\mathrm{We}_\mathrm{local}=0.37$. These are very small values of the Weber number, which likely indicates that aerodynamic forces are not major factors in breakup in this system. This is logical, because the simulation is liquid injected into quiescent air.

\subsection{Atomization Evolution}
The evolution of local droplet characteristics and global jet development throughout the atomization process are important attributes, which can help elucidate the underlying physics of atomization. Understanding how droplets change from first breakup off of the liquid core to a final droplet in dilute spray could provide useful information to atomization model developers seeking to not only describe the final spray formation, but also the intermediate spray development. The tool provided in this work enables researchers to extract droplet statistics from throughout a simulation and analyze the evolution of these droplets and the system as a whole. 

\subsubsection{Droplet Shape and Size Evolution}
\begin{figure}[!htbp]
 \centering
 \includegraphics[width=1\textwidth]{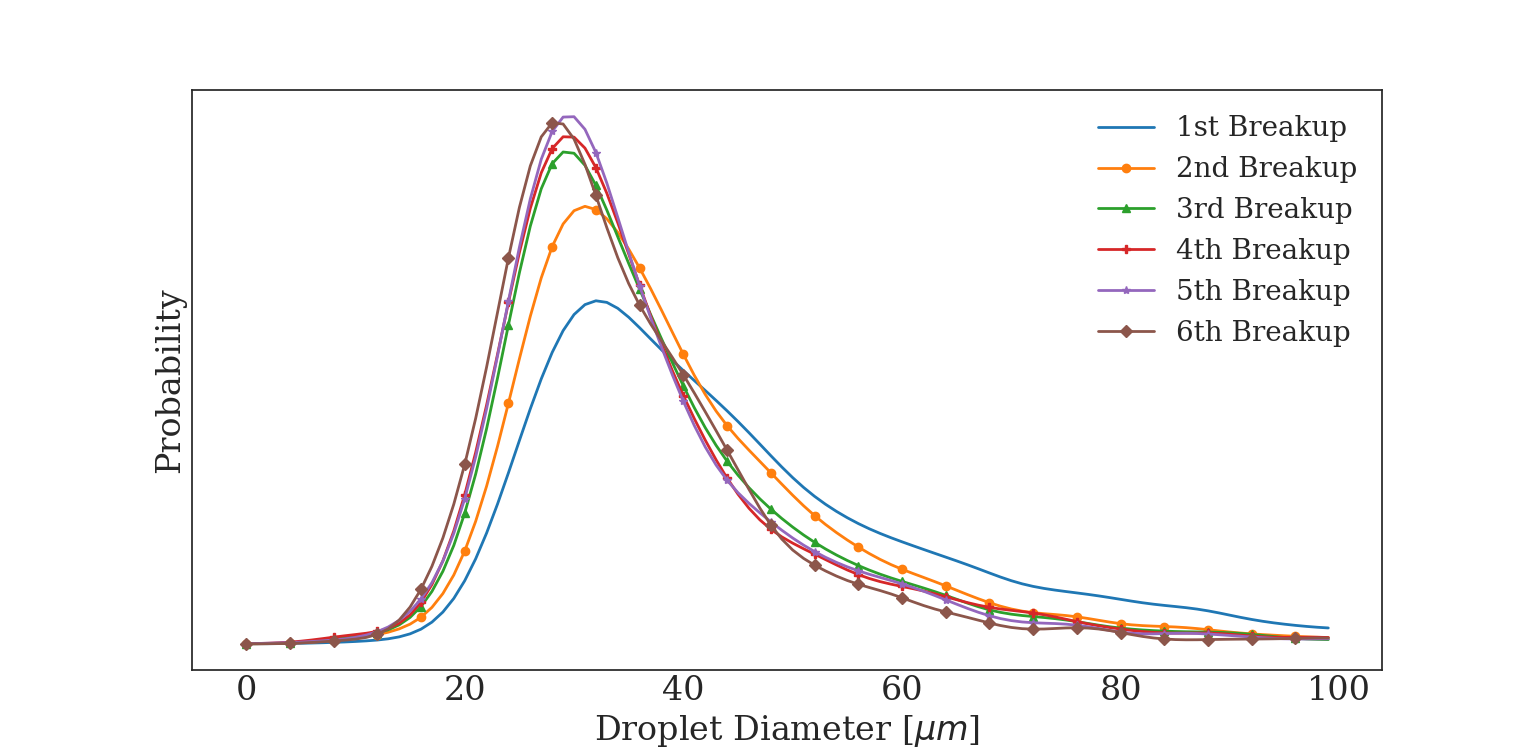}
 \caption{A probability density function of droplet diameter as a function of number of breakups} 
 \label{fig:diam_pdf}
\end{figure}

\begin{figure}[!htbp]
 \centering
 \includegraphics[width=1\textwidth]{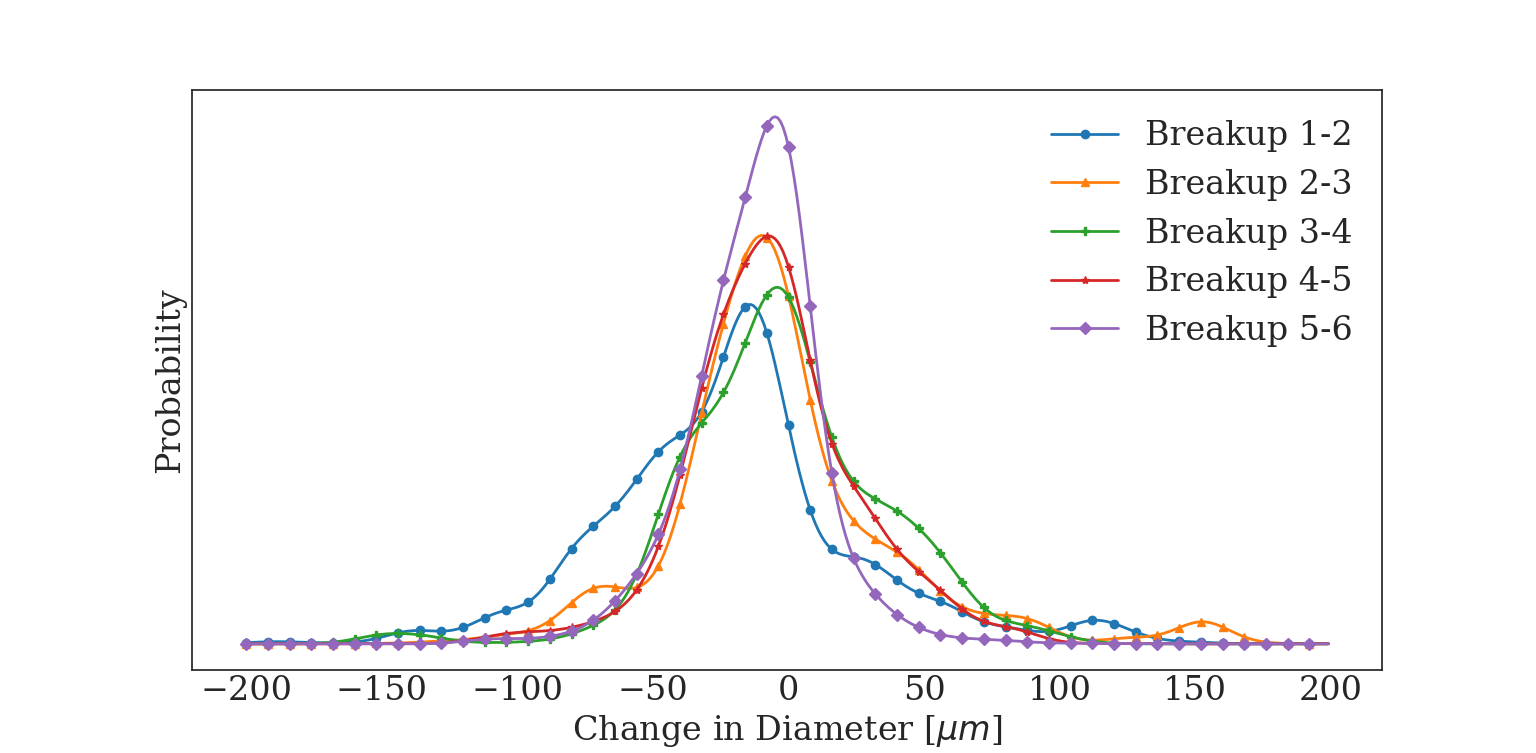}
 \caption{A probability density function displaying the change in diameter between breakup events.} 
 \label{fig:diam_change}
\end{figure}

Fig.~\ref{fig:diam_pdf} displays a probability density function of the equivalent spherical diameter of droplets as a function of the number of breakup events. A reasonable trend toward smaller and more uniformly sized droplets is seen, which provides confidence in the values extracted by the tool. 
Fig.~\ref{fig:diam_change} displays the change in diameter of droplets between breakup stages. Notice that not all values are negative. This further confirms the prevalence of coalescence events within these systems. Positive values in the figure indicate that a significant portion of droplets undergo coalescence and increase in size between breakup events. Additionally, notice the tendency of droplets to change size less as more breakup events occur, with the PDFs becoming more narrow and centered around 0 $\mu$m.

\subsubsection{Time Evolution} \label{sec:time}
\begin{figure}[!htbp]
 \centering
 \includegraphics[width=1\textwidth]{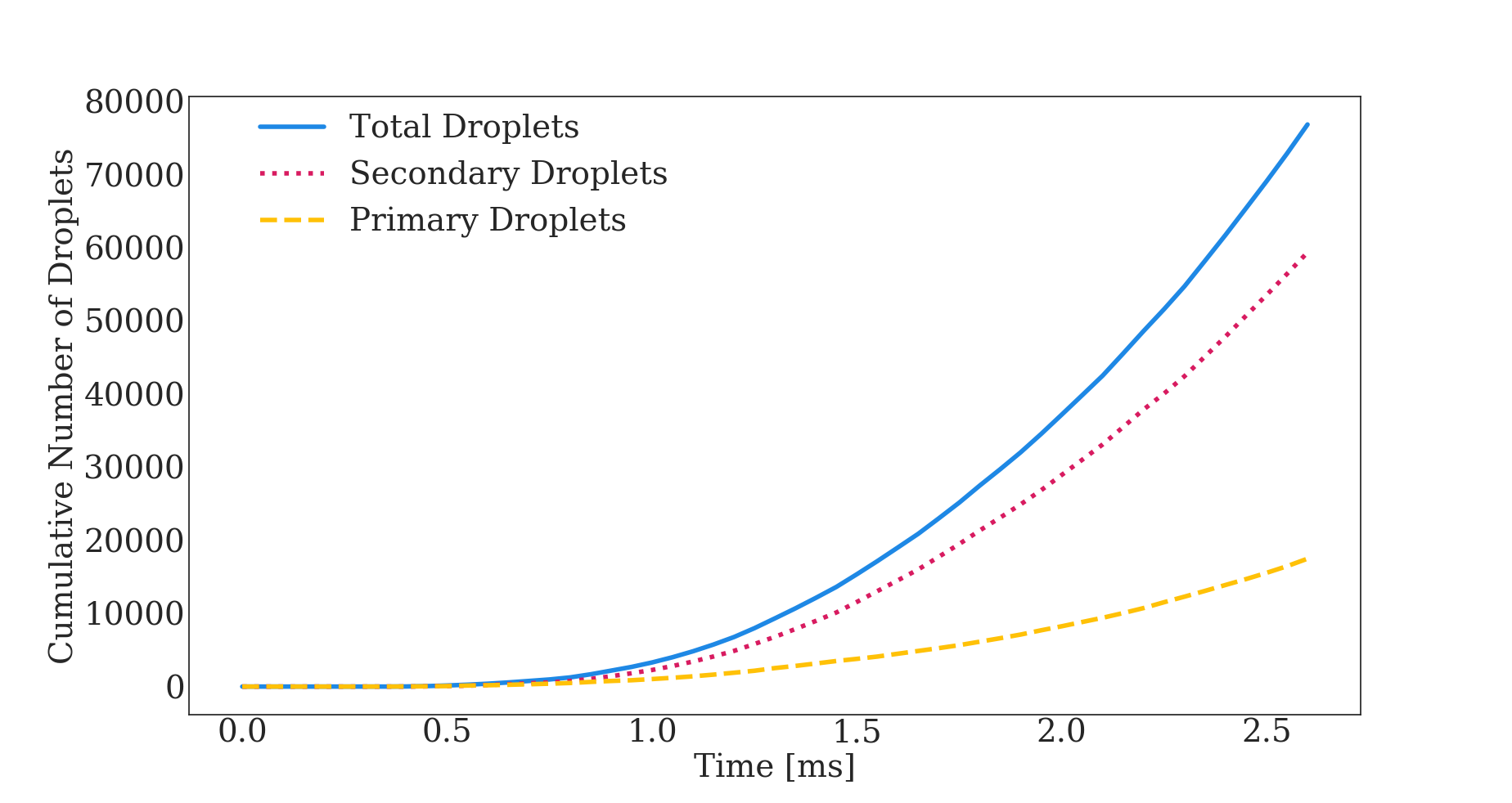}
 \caption{The cumulative number of primary, secondary, and total droplets. Note: secondary droplets in this image represent the classical definition of secondary droplets (i.e. all breakup following primary).} 
 \label{fig:drop_vol_evo}
\end{figure}

Because the extraction tool samples from every timestep in which breakup occurs. Thus, once the jet develops, breakup occurs on every timestep, providing high resolution time-series statistics describing evolution of the jet. Fig.~\ref{fig:drop_vol_evo} shows an example of some of the time dependent statistics that were extracted in the present work. These plots describe the rate of droplet development in the simulation. It is clear from this plot that secondary atomization quickly dominates droplet production in this system. Future work will focus on improving the extracted quantities and further explore how atomizing systems develop through time. 

\section{Conclusions}
An extraction tool for numerical simulations of atomization was developed and tested. Improvements upon the methodology originally proposed by Rubel and Owkes~\cite{RubelOwkes2019} were made. Improvements include 1) an updated algorithm to stop fictitious identification of merge and split events, 2) enabling a wider temporal sampling range, and 3) the extraction of more atomization statistics.

A diesel-type jet was simulated and droplet statistics were extracted. The tool displayed utility for extracting data relevant to atomization model developers as well as providing previously inaccessible information on the underlying mechanisms of atomization. The data extracted are preliminary, but show promise for the tool's utility in future high-resolution studies. 

Future work will focus on the implementation of the tool into high-resolution atomization simulations to extract useful data to help better quantify atomization mechanisms on global and local scales. Special effort will be given to aiding reduced-order model developers to develop relevant statistics for creating more accurate models. Additionally, we will focus on the improvement of sampling methods, particularly the local flow field to attempt to extract not only velocity magnitudes, but also topological flow data to better understand the small-scale turbulence that effects liquid breakup. 

\section{Acknowledgements}
This material is based upon work supported by the National Science Foundation under Grant No.~1749779.
\vskip 4mm
Computational efforts were performed on the Hyalite High Performance Computing System, operated and supported by  University Information Technology Research Cyberinfrastructure at Montana State University.

\bibliographystyle{unsrt}
\bibliography{CF_Paper.bib}
\appendix
\newpage
\section{Liquid Identification and Coalescence Code}\label{appendixa}

    The following is a pseudocode describing how the liquid identification number is transported and coalescence is identified. 

    \begin{algorithm}[H]
    \caption{$\mathcal{L}$-transport and Coalescence Pseudocode}\label{alg:cap}
    \begin{algorithmic}[1]
    \For{$i = 1 \rightarrow N_\mathrm{cell}$} \Comment{Loop over cells in domain}
        \State \textbf{$\triangleright$ Form list of $\mathcal{L}$ for this cell}
        \State $\mathcal{L}_\mathrm{count}=0$
        \If{cell has interface or VoF > 0.5} \Comment{Search through cells with liquid}
            \If{$\mathcal{L}$ > 0}
                \State $\mathcal{L}_\mathrm{count} \gets \mathcal{L}_\mathrm{count} + 1$ \Comment{Keep track of the number of unique $\mathcal{L}$ values within the cell}
                \State $\mathcal{L}_\mathrm{cell}(\mathcal{L}_\mathrm{count}) \gets \mathcal{L}$ \Comment{Keep track of $\mathcal{L}$ value in the cell}
            \EndIf
        \EndIf 
        \For{$f=1\rightarrow N_\mathrm{faces}$} \Comment{ Loop over all faces of a cell} 
            \For{$n = 1 \rightarrow \mathcal{L}_\mathrm{max}$} \Comment{Loop through all $\mathcal{L}$ that can flux through a face}
                \If{$\mathcal{L}_{\mathrm{flux},n} \neq 0$} \Comment{Check to see if there is an $n^\mathrm{th}$ $\mathcal{L}$ flux into the cell}     
                    \If{$\mathcal{L}_{\mathrm{flux},n} \neq \mathcal{L}_\mathrm{cell}(1:\mathcal{L}_\mathrm{cell})$} \Comment{Check to see if $\mathcal{L}$ fluxes matches any of the current $\mathcal{L}$ in the cell}
                        \State $\mathcal{L}_\mathrm{count} \gets \mathcal{L}_\mathrm{count} + 1$ \Comment{Update $\mathcal{L}$ counter in the cell}
                        \State $\mathcal{L}_\mathrm{cell}(\mathcal{L}_\mathrm{count}) \gets \mathcal{L}_\mathrm{flux}$ \Comment{Keep track of all $\mathcal{L}$ values in the cell}
                    \EndIf
                \EndIf
            \EndFor
        \EndFor
        \State \textbf{$\triangleright$ Check for coalescence or merge event}
        \If{$\mathcal{L}_\mathrm{count} \geq 2$} \Comment{More than one unique $\mathcal{L}$ for this cell}
            \For{$n = 2 \rightarrow \mathcal{L}_\mathrm{count}$}
                \State nMerge = nMerge+1 \Comment{Keep track of the number of coalescence events}
                \State $\mathcal{L}1$(nMerge) = min($\mathcal{L}_\mathrm{cell}(1),\mathcal{L}_\mathrm{cell}(n)$) \Comment{Make a list of the $\mathcal{L}$'s which need to be merged}
                \State $\mathcal{L}2$(nMerge) = max($\mathcal{L}_\mathrm{cell}(1),\mathcal{L}_\mathrm{cell}(n)$)
            \EndFor
        \EndIf
    \EndFor
    \State \textbf{$\triangleright$ Compute new $\mathcal{L}$ after coalescence event}
    \For{$n = 1 \rightarrow$ nMerge} \Comment{Loop over the all identified coalescence events}
        \If{$\mathcal{L}1(n)$ = $\mathcal{L}1(n+1)$}\Comment{Combine identical events}
            \If{$\mathcal{L}2(n)$ = $\mathcal{L}2(n+1)$} 
                \State $\mathcal{L}_{new} \gets min(\mathcal{L}1,\mathcal{L}2$) \Comment{Assign the smaller $\mathcal{L}$ to the new merged structure}
                \State $\mathcal{L}_{old} \gets max(\mathcal{L}1,\mathcal{L}2$)
            \EndIf
        \EndIf
    \EndFor
    \State \textbf{$\triangleright$ Update $\mathcal{L}$ in domain}  
    \State \textbf{where} ($\mathcal{L} = \mathcal{L}_{old}$): $\mathcal{L} \gets \mathcal{L}_{new}$ \Comment{Update $\mathcal{L}$'s throughout domain}
    
    \end{algorithmic}
    \end{algorithm}
\newpage
\section{Neo4j Data Input}\label{appendixb}
\lstset{
	language=SQL, 
	morekeywords={RETURN,SHORTESTPATH,LENGTH,LOAD,CSV,
	WITH,HEADERS,TOINTEGER,TOFLOAT,CALL,YIELD},
	deletekeywords={PRIMARY},
	breaklines=true,  
	breakatwhitespace=false,   
	morecomment=[f][\color{gray}][0]{//}, 
}

Data from our simulation are exported in the form of a CSV file. Each row within the CSV represents a breakup event within the simulation and each column contains statistics we extracted from the event. Column headers are listed in the following Cypher code, following the "csvline" variable (i.e. csvline.OldLID and csvline.NewLID are the old and new $\mathcal{L}$'s associated with each droplet). We import the CSV into Neo4j to create nodes and relationships. Each node corresponds to a row within the CSV and the relationships are created using the identification numbers. 

\begin{lstlisting}
// Import data from CSV and create split/merge nodes
LOAD CSV WITH HEADERS FROM "link_to_csv" AS csvline CREATE (n:droplet {id: TOINTEGER(csvline.NewLID), Event: csvline.Merge_Split, OldLID: TOINTEGER(csvline.OldLID), OldSID: TOINTEGER(csvline.OldSID), NewLID: TOINTEGER(csvline.NewLID), NewSID: TOINTEGER(csvline.NewSID), Volume: TOFLOAT(csvline.Vol), Event_Time: TOFLOAT(csvline.Time), X:  TOFLOAT(csvline.X), Y:  TOFLOAT(csvline.Y), Z:  TOFLOAT(csvline.Z), U:  TOFLOAT(csvline.U), V:  TOFLOAT(csvline.V), W:  TOFLOAT(csvline.W), U_gas:  TOFLOAT(csvline.U_gas), V_gas:  TOFLOAT(csvline.V_gas), W_gas:  TOFLOAT(csvline.W_gas), L1_old:  TOFLOAT(csvline.L1_old), L2_old:  TOFLOAT(csvline.L2_old), L3_old:  TOFLOAT(csvline.L3_old), L1_new:  TOFLOAT(csvline.L1_new), L2_new:  TOFLOAT(csvline.L2_new), L3_new:  TOFLOAT(csvline.L3_new), Vol_old: TOFLOAT(csvline.Vol_old)})

// Create merge relations between droplets
MATCH (n:droplet{Event:"Split"}),(d:droplet{Event:"Split"}),(m:droplet{Event:"Merge"})
WHERE n.NewLID = m.NewLID AND d.NewLID = m.OldLID
CREATE (d)-[:Merge]->(n)

// Create merge relations between droplets and core
MATCH (n:droplet{Event:"None"}),(d:droplet{Event:"Split"}),(m:droplet{Event:"Merge"})
WHERE n.NewLID = m.NewLID AND d.NewLID = m.OldLID
CREATE (d)-[:Merge]->(n)

// Deletes merge nodes
MATCH (d:droplet)
WHERE d.Event = 'Merge'
DELETE d;

// Creates split relations
MATCH (n:droplet),(d:droplet)
WHERE n.Event = "Split" and n.OldLID = d.NewLID
CREATE (d)-[:Split]->(n);
\end{lstlisting}

\section{Python Neo4j Driver Script}
The below script uses the py2neo library to query a Neo4j graph database from Python. This script was used to analyze the data presented in Fig.~\ref{fig:drop_vol_evo}. First, communication with a graph is established. The graph must be running in Neo4j when the script is executed. Then, a time series is created and result arrays are initialized. Following this, the Cypher queries are executed. To use a Python variable within the Cypher query, a \$ must be placed immediately before the variable, then the variable is defined in the second argument in the ".run" function. The ".evaluate()" function ensures that only values from the query are returned, otherwise results appear as a dictionary. Results from the query are then appended to the results lists to be used for further analysis. 

\begin{lstlisting}[language=Python]
import numpy as np
from py2neo import Graph
# Using Python Neo4j API "py2neo" to query a graph database
gs = Graph("<Graph URI>",password="<Graph Password>") # Graph must be running on the Neo4j platform for this to work. First input is graph URI and second is a required password

# Set up time array
dt = 5e-5
time = []
x = np.arange(0,2.65e-3,dt)

# Convert time class from numpy.float64 to python native float
for t in x:
    time.append(t.item())

# Initialize result arrays
nd = [] # total number of droplets
n1 = [] # number of primary droplets
n2 = [] # total number of secondary droplets

# Run Cypher Queries

# Total Droplets
for t in time:
    g = gs.run("MATCH(n:droplet),(c:core) WHERE n.Event_Time <= $t RETURN count(n)",t=t).evaluate() # evaluate() returns only values from the query
    nd.append(g)
# Primary Droplets
for t in time:
    g = gs.run("MATCH(n:primary),(c:core) WHERE n.Event_Time <= $t RETURN count(n)",t=t).evaluate()
    n1.append(g)
# All secondary
for t in time:
    g = gs.run("MATCH(n:droplet),(c:core) WHERE n.Event_Time <= $t AND NOT n:primary RETURN count(n)",t=t).evaluate()
    n2.append(g)
\end{lstlisting}

\end{document}